\documentclass{appolb}
\usepackage{amsmath,amsfonts,amssymb}
\usepackage{relsize}
\usepackage[dvips]{graphicx}

\newcommand{\be}{\begin{equation}}
\newcommand{\ee}{\end{equation}}
\newcommand{\ba}{\begin{array}}
\newcommand{\ea}{\end{array}}
\newcommand{\bea}{\begin{eqnarray}}
\newcommand{\eea}{\end{eqnarray}}

\newcommand{\bi}{\begin{itemize}}
\newcommand{\ei}{\end{itemize}}
\newcommand{\bal}{\begin{aligned}}
\newcommand{\eal}{\end{aligned}}

\begin{document}
\title{Deconfinement temperature in AdS/QCD from the spectrum of scalar glueballs\thanks{Presented at "Excited QCD 2019", Schladming, Austria, Jan.30  –- Feb.3, 2019.}}
\author{S. S. Afonin, A. D. Katanaeva, E. V. Prokhvatilov and M. I. Vyazovsky\address{Saint Petersburg State University, 7/9 Universitetskaya nab., St.Petersburg, 199034, Russia}}
\maketitle

\begin{abstract}
We scrutinize various holographic estimations of the deconfinement temperature within the bottom-up AdS/QCD models. A special emphasis is put on the recent idea of isospectral potentials in the holographic approach. It is demonstrated that different models from an isospectral family (i.e., the models leading to identical predictions for the spectrum of hadrons with fixed quantum numbers) result in different predictions for the deconfinement temperature. This difference is found to be quite small in the scalar glueball channel but very large in the vector meson channel which is often used for fixing parameters of holographic models. The observed stability in the former case clearly favors the choice of the glueball channel for thermodynamic predictions in AdS/QCD models, with the scalar glueball trajectory being taken from lattice simulations and used as a basic input in improved versions of the Soft Wall holographic model.
\end{abstract}
\PACS{12.40.Yx, 12.39.Mk}

\bigskip
\bigskip

The modern experiments on heavy ion collisions at ALICE in CERN and RHIC (the Brookhaven National
Laboratory) and planned experiments at FAIR (GSI) have caused a hot interest in the theoretical study of the QCD phase
diagram.
Some time ago it was realized that the QCD phase diagram can be studied within the framework of the holographic approach to strong interactions.
The holographic approach has many applications in QCD~\cite{add1} and beyond, including even a description of high temperature superconductivity
(see, e.g., discussions in~\cite{add10}). Concerning the QCD phase diagram,
one of the primary questions is to calculate the critical
temperature $T_c$ at which hadronic matter is supposed to undergo
a transition to a deconfined phase.
Within the bottom-up holographic approach this type of studies was pioneered by Herzog in Ref.~\cite{Herzog:2006} and continued by
many authors (see, e.g., references in \cite{estim_deconf}).
In these papers the gravity part of a $5D$ model is assumed to come from a dual description of
gluodynamics and can be used to study thermodynamic properties of original $4D$ gauge theory.
The deconfinement is related to a Hawking--Page phase transition
between a low temperature thermal Anti-de Sitter (AdS) space and a high temperature
black hole in the AdS/QCD models.

The two sectors of the $5D$ model turn out to be closely related when concerning the value of $T_c$.
The estimation of of $T_c$ depends on the model parameters which were traditionally connected to the main
purpose of the first bottom-up AdS/QCD papers (\cite{HW_2005, HW2, SW_2006})
-- to the description of light vector meson spectra (see, e.g., discussions in~\cite{add1,add2,add3,add4,add5}). The estimates of $T_c$ in \cite{Herzog:2006}
follow the parameter values of these works.
This traditional choice was motivated by a relatively rich experimental data on light vector mesons, however,
there seems to be no particular reason why the vector meson spectra should
in any way determine the deconfinement temperature. In the same time, a simple description of meson spectra, especially the radial Regge
trajectories appearing in Soft Wall (SW) models~\cite{SW_2006}, are one of the most attractive features of holography.

In the present study, we will argue that the scalar glueball (and its radial excitations) is much better option for
fixation of the model parameters.
Our main arguments can be shortly given as:\\
(i) Phase diagram can be studied in pure gluodynamics.
 Since the holographic approach is defined in the large-$N_c$ (planar) limit of gauge theories
 where the glueballs dominate over the usual mesons and baryons (as the quarks are in the fundamental
 representation) the gluodynamics must dictate the overall mass scale and thereby the major contribution
 to the deconfinement temperature $T_c$.\\
(ii) The isospectrality concept~\cite{Vega_Cabrera}. Within it one can show that the predicted values of $T_c$ are more stable
for scalar glueballs than for vector mesons.\\
(iii) Phenomenological reasons. Numerical values of $T_c$ determined in the scalar glueball framework can be interpreted as better fitting the lattice expectations.

The first argument was scrutinized in our paper \cite{estim_deconf} (see also~\cite{add6,add7}) and we will further substantiate the point. One can observe, for example,
that if we take the linear radial spectrum of scalar glueballs given by the standard SW holographic model,
$m_n^2=\mu^2(n+2)$, $n=0,1,2,\dots$  (the interpolating operator $G_{\mu\nu}^2$ is assumed), and consider the scalar resonance
$f_0(1500)$ \cite{PDG} as the lightest glueball (as is often proposed in the hadron spectroscopy \cite{PDG}) we will obtain the slope $\mu^2=1.13$ GeV$^2$.
It agrees perfectly with the mean radial slope $\mu^2=1.14\pm 0.01$ GeV$^2$ found for the light mesons in the analysis of Ref.~\cite{bugg} and achieved independently
in Ref.~\cite{Afonin:unflmeson} (see also some related discussions in~\cite{add8}).

The second argument is motivated by an interesting finding of the authors of Ref.~\cite{Vega_Cabrera} that the particular form
of spectrum in SW-like models does not fix a model itself:
One can always construct a one-parametric family of SW models (controlling modifications of the "wall") leading to the same spectrum. We will demonstrate that such isospectral models, however,
result in different predictions for the deconfinement temperature $T_c$. In the vector channel these variations can
be quite significant, whereas in the scalar case the difference may be rather small and spanning
an interval admissible by the accuracy of the large $N_c$ limit.

The third argument is just a phenomenological observation for typical predictions of $T_c$ in the bottom-up holographic models. For instance, the original Herzog's
analysis of the vector Hard Wall (HW) holographic model~\cite{HW_2005} with the $\rho$-meson taken as the lowest state resulted in the prediction $T_c=122$ MeV~\cite{Herzog:2006}.
If we apply this analysis to the scalar HW model with $f_0(1500)$ as the lightest glueball, we will find $T_c \approx 150$ MeV.
The lattice simulations typically predict the lightest glueball near $1.7$ GeV ($1.6 - 1.7$ GeV as quoted in Particle Data reviews \cite{PDG}). This value shifts the prediction to $T_c \approx 170$ MeV.
So we may regard the interval $T_c = 150 - 170$ MeV as a prediction of the HW model in the glueball channel.
We find remarkable that exactly this interval was found in the modern lattice simulations  with dynamical
quarks \cite{Borsanyi:2010bp}.

We will consider scalar glueballs and vector mesons in parallel
in order to demonstrate in detail the emergent differences.

Let us introduce the $5D$ holographic action with a universal gravitational part and a matter part to be specified,
it has an AdS related metric $g_{MN}$ ($g=\det g_{MN}$)
\bea
S &=& \int d^4xdz\sqrt{-g} f^2(z)\left(\mathcal L_{gravity} + \mathcal L_{matter} \right),\\
\mathcal L_{gravity} &=& -\frac1{2k_g} \left(\mathcal{R}-2\Lambda\right).
\eea
Here $k_g$ is the coefficient proportional to the $5D$ Newton constant, $\mathcal{R}$ is the Ricci scalar and $\Lambda$ is the cosmological constant.
The choice of the dilaton background, $f(z)$, distinguishes possible holographic models.
They differ as well by the interval the $z$ coordinate spans. For now we assume $z\in [0,z_{max}]$,
though $z_{max}=\infty$ is possible and will be of the main interest in the present work as it corresponds to the SW model.

The assessment of the critical temperature is related to the leading contribution in the large $N_c$ counting,
that is the $\mathcal L_{gravity}$ part scaling as $\frac1{2k_g} \sim N_c^2$ while $\mathcal L_{matter}$ scales
as $N_c$.
According to \cite{Herzog:2006} the deconfinement in AdS/QCD occurs as the Hawking -- Page phase transition that is a transition between different gravitational backgrounds.
We call the order parameter of this transition $\Delta V$.

$V$'s are the free action densities evaluated on different backgrounds corresponding to two phases.
First, the confined phase.
It is given by the thermal AdS of radius $R$ and defined by the general AdS line element
\be
ds_{Th}^2=\frac{R^2}{z^2}\left(dt^2-d\vec{x}^2-dz^2\right),
\ee
with the time direction restrained to the interval $[0, \beta]$.
The metric of the Schwarzschild black hole in AdS describes the deconfined phase and is given by
\be
ds_{BH}^2=\frac{R^2}{z^2}\left(h(z)dt^2-d\vec{x}^2-\frac{dz^2}{h(z)}\right),
\ee
where $h(z)=1-(z/z_h)^4$ and $z_h$ denotes the horizon of the black hole.
The corresponding Hawking temperature is related to the horizon as $T=1/(\pi z_h)$.

The cosmological constant in $5D$ AdS is $\Lambda=-6/R^2$ and
both these metrics are the solutions of the Einstein equations. They provide the same value of the Ricci scalar $\mathcal R = -20/R^2$.
Hence, the free action densities differ only in the integration limits,
\begin{align}
V_{\text{Th}}(\epsilon)&=\frac{4R^3}{k_g} \int_0^\beta dt\int_\epsilon^{z_{max}}dzf^2(z) z^{-5},\\
V_{\text{BH}}(\epsilon)&= \frac{4R^3}{k_g}  \int_0^{\pi z_h}dt\int_\epsilon^{\min(z_{max},z_h)}dzf^2(z)z^{-5}.
\end{align}
The two geometries are compared at $z=\epsilon$ where the periodicity in the
time direction is locally the same, {\it i.e.} $\beta=\pi
z_h\sqrt{h(\epsilon)}$. Then, we may construct the order parameter for the phase
transition,
\begin{equation}
\label{deltaV}
\Delta V = \lim_{\epsilon\rightarrow0}\left(V_{\text{BH}}(\epsilon)-V_{\text{Th}}(\epsilon)\right).
\end{equation}
The thermal AdS is stable when $\Delta V>0$, otherwise the
black hole is stable. The condition $\Delta V=0$ defines the critical temperature $T_c$ at which the
transition between the two phases happens.
Eqn.~(\ref{deltaV}) yields $z_h$ as a function of the model dependent parameters -- $z_{max}$ and/or those possibly
introduced in $f(z)$. We must appeal to the matter sector $\mathcal L_{matter}$ to give physical meaning to
these parameters and to connect $T_c$ to a particular type of a holographic model.

As was recently noticed in Ref.~\cite{Vega_Cabrera}, the SW background is not fixed by the form of linear spectrum as
one can find an infinite number of one-dimensional potentials leading to identical spectrum of normalized modes.
The corresponding family of potentials is referred to as isospectral potentials.

In brief, the problem for mass spectrum in the bottom-up holographic models can be reduced to a one-dimensional Schr\"{o}dinger equation
\be
\label{Schrod}
-\psi''_n(z)+\widehat{\mathcal V}(z) \psi_n(z) =M^2(n)\psi_n(z),
\ee
where $\widehat{\mathcal V} (z)$ is the Schr\"{o}dinger potential which depends on the 5D dilaton background
$f^2(z)$, metric, and spin.
A particular form of the Schr\"{o}dinger potential defines the eigenvalues of Eqn.~(\ref{Schrod}) and hence
the mass spectrum $M(n)$. In the case of SW models it is a potential similar to the one that appears
when considering the radial part of the wavefunction of a $2D$ harmonic oscillator system.

According to~\cite{Vega_Cabrera} and the references therein, there is
the following isospectral transformation between $\mathcal V_J(z)$ and  $\widehat{\mathcal V}_J(z)$,
\be
\widehat{\mathcal V}_J(z) = \mathcal V_J(z) -2\frac{d^2}{dz^2}\ln[I_J(z)+\lambda].
\ee
This technique allows one to generate a family of dilaton functions $f(z)$ appearing in $\widehat{\mathcal V}_J(z)$,
each member assigned to the value of the parameter $\lambda$. The case of $\lambda=\infty$ corresponds to the original $\mathcal V_J(z)$.
The function $I_J(z)$ is defined through the ground eigenstate of $\mathcal V_J$, $\psi_0$, and is given by
\be
I_J(z)\equiv\int\limits_0^z\psi_0^2(z')dz'.
\ee
Different $\lambda$ provide slightly different form of the potential, but the eigenvalues of Eqn.~(\ref{Schrod})
and, hence, the spectrum remain the same.

The main problem we studied can be formulated as follows: Does isospectrality entail isothermality (i.e. identical predictions for $T_c$)?
In general case, the answer turns out to be negative. But there is one important exception: If model parameters are fixed from the
scalar glueball channel within the generalized SW holographic model of Ref.~\cite{GSW,add9} (which is able to reproduce accurately the glueball
radial spectrum) then an isospectral family of models leads to almost identical predictions for the deconfinement temperature. The
typical predictions lie in the range $T_c\simeq 175 \pm 15$~MeV which agrees very well with modern unquenched lattice estimations.
The details of further continuation of the reported analysis are contained in Ref.~\cite{AK2018}.

\section*{Acknowledgements}

The present work was supported by the Saint-Petersburg State University
travel grant (Id: 36042146).

\end{document}